# Unusual Compression Behavior of Columbite $TiO_2$ via First-Principles Calculations

**Xiang-Feng Zhou[1], Xiao Dong[1], Guang-Rui Qian[3], Lixin Zhang[1], Yonjun Tian[2], and Hui-Tian Wang[1,3]**

[1]School of Physics and Key Laboratory of Weak-Light Nonlinear Photonics, Ministry of Education, Nankai University, Tianjin 300071, China

[2]State Key Laboratory of Metastable Materials Science and Technology, Yanshan University, Qinhuangdao 066004, China

[3]Nanjing National Laboratory of Microstructures, Nanjing University, Nanjing 210093, China

## Abstract

The physical mechanisms behind the reduction of the bulk modulus of a high-pressure cubic $TiO_2$ phase are confirmed by first-principles calculations. An unusual and abrupt change occurs in the dependence of energy on pressure at 43 GPa, indicating a pressure-induced phase transition from columbite $TiO_2$ to a newly-identified modified fluorite $TiO_2$ with a Pca21 symmetry. Oxygen atom displacement in Pca21 $TiO_2$ unexpectedly reduces the bulk modulus by 34% relative to fluorite $TiO_2$. This discovering provides a direct evidence for understanding the compressive properties of such groups of homologous materials

Titanium dioxide ($TiO_2$) has rich phase diagrams, namely, the rutile (P42/mnm), anatase (I41/amd), brookite (Pbca), columbite (Pbcn), baddeleyite (P21/c), and cotunnite (Pnma) phases [1-6]. Due to its versatile physical and chemical properties, $TiO_2$ is extensively used in many industrial applications, such as high efficiency solar cells, photocatalysis, dynamic random access memory modules, and super-hard materials [7-12]. The rutile and anatase phases of $TiO_2$ are abundant in nature [13,14]. Since the phase sequence of $TiO_2$ is very similar to that of other bulk materials, such as $ZrO_2$ and $HfO_2$, it is highly expected to transform into its cubic polymorphs under pressure [15]. Modified cubic fluorite-structured $RuO_2$, $SnO_2$, and $PbO_2$ that possess a Pa-3 symmetry, have been successfully synthesized [16]. In particular, $RuO_2$ is considered to be a potential ultra-hard material because of its measured Knoop hardness (~20 GPa) and bulk modulus (399 GPa), which is only 10% less than that of sintered diamonds [17]. Moreover, synthesized cotunnite $TiO_2$ has an extremely high bulk modulus of 431 GPa and is considered as the hardest oxide to date [1]. After the synthesis of cotunnite $TiO_2$, scientists expected to synthesize cubic $TiO_2$ because it showed potential for use as a solar cell or ultra-hard material. Ultimately, the highly anticipated cubic $TiO_2$ was successfully synthesized by heating anatase $TiO_2$ between 1900 and 2100 K in diamond-anvil cells under a pressure of 48 GPa [18]. Some ambiguities, however, remained both in the experiment and theory. For instance, the theoretical bulk modulus calculated for cubic $TiO_2$ in the pyrite and fluorite phases was significantly larger than that obtained during the experiments. Kim *et al.* showed that pyrite $TiO_2$ is unstable because of the presence of imaginary frequencies in the phonon spectra throughout the entire pressure range, whereas fluorite $TiO_2$ is stable because of the absence of these imaginary frequencies under pressure [13]. Swamy and Muddle reported that pyrite $TiO_2$ has theoretical properties closer to the experimental values, because it has a relatively lower bulk modulus [19]. In terms of mechanical properties, however, Liang *et al.* [20] found a minor difference between the fluorite and pyrite phases. They found that the fluorite $TiO_2$ are closer to the experimental one when calculation is done using different codes. This was opposite to the claim made in Ref. [19]. Consequently, even though many efforts have been made to elucidate its properties, some features of cubic $TiO_2$ remain questionable.

In this work, first-principles calculations are performed using the projector augmented wave method implemented in the *ab initio* total and molecular-dynamics program, VASP [21]. We employ a generalized gradient approximation for the exchange correlation functional and used a cut-off energy of 500 eV and a Monkhorst-Pack Brillouin zone sampling grid spacing of 0.5 $Å^{-1}$. During the geometry optimization process, no symmetry and no restrictions are constrained for both the unit cell and the atomic positions. A residual minimization scheme and direct inversions in the iterative subspace are employed. Structural relaxation is prevented until the total energy is less than $10^{-5}$ eV and the force is less than $10^{-2}$ eV/ Å. A $2\times2\times2$ supercell is used to perform phonon dispersion calculations using the PHONON code within the *ab initio* force-constant method. [21] Powder x-ray diffraction (XRD) patterns are simulated by the REFLEX software [22].

We begin from the columbite structure and impose hydrostatic pressure on it. As the pressure increases from ambient conditions, we obtain a series of optimized configurations after relaxing the structure under pre-defined pressure points. The calculated pressure dependency of energy exhibits an unusual and abrupt change at a pressure of 43 GPa, suggesting the occurrence of structural transition and the appearance of a new phase. By analyzing the symmetry of this phase, the new phase is determined to be that of a modified fluorite structure.

By carefully examining the trajectory file at a transition pressure of 43 GPa, we found that the evolution process of the phase transition can be divided into three stages, as shown by the circles in Fig. 1: (i) The symmetry retains the original symmetry (Pbcn) of columbite $TiO_2$ from the 1st step to the 30th step. (ii) The columbite $TiO_2$ transforms to a modified columbite structure from the 31st step to the 40th step. (iii) Finally, the modified columbite $TiO_2$ transforms into the modified fluorite $TiO_2$ at the 41st step. Figure 2 shows the projections of the structures along the [010] direction of a $2\times2\times2$ columbite $TiO_2$ supercell at a transition pressure of 43 GPa. Figures 2(a) and 2(b) show snapshots of the 1st and 30th steps. Figure 2(c) shows the final structure. For comparison, the fluorite structure is also shown in Fig. 2(d). Clearly, the O atoms undergo large displacements, and the Ti atoms exhibit smaller ones [Figs. 2(a)-(c)]. As shown in Fig. 2(a), four nearest-neighbor Ti atoms exhibit a near rhombic motif in the starting columbite phase. In contrast, as shown in Fig. 2(c), four nearest-neighbor Ti atoms form a square motif. The O atom in these figures exhibits significant deviation from that in fluorite $TiO_2$ in Fig. 2(d). Therefore, we became interested in the final structure of $TiO_2$ and the factors that influence the mechanical properties of the high pressure phase of cubic $TiO_2$. The modified fluorite $TiO_2$ with a tolerance of 0.5 Å has a Fm-3m symmetry, which is the same as that of the fluorite $TiO_2$. When the tolerances are 0.1, 0.01, and 0.001 Å, however, the $TiO_2$ structure is not fluorite; rather, the resulting structures have P42/nmc, Aba2, and Pca21 symmetries. This is determined using *Find Symmetry* technology [23]. The multifold symmetries of the modified fluorite $TiO_2$ originate from the uncertainty of the O atom positions. This is responsible for the relatively unstable bonding in the modified fluorite $TiO_2$ compared with the fluorite $TiO_2$.

To explore the influence of the O atom displacements, we simulated the XRD patterns of Pca21 $TiO_2$, fluorite $TiO_2$, and pyrite $TiO_2$ and compared them with the experimental data. Mattesini *et al*. claimed that the fluorite and distorted fluorite phases (Pa-3) cannot be unambiguously distinguished because some weak XRD peaks are screened by the XRD peaks of the cotunnite phase [18]. Our simulation results in Fig. 3 clearly show that the differences in the O positions of pyrite, fluorite, and Pca21 phases significantly influence both the positions and relative intensities of the peaks in the XRD patterns. The calculated displacements of the O atoms in the Pca21 $TiO_2$ phase match the experimental results more closely than the other two phases [18]. In particular, the intensity ratios of the 220 peak to the 111 peak are 57% for the pyrite phase, 99% for the fluorite phase, and 33% for the Pca21 phase. The experimental value is 45%. The intensity ratios of the various 113 peaks to 111 peaks are 46% for the pyrite phase, 84% for the fluorite phase, and 24% for the Pca21 phase. The experimental value is 31%. The residual weak peaks, including the one at 200, also more closely match the experimental data. Consequently, the Pca21 $TiO_2$ phase has the closest match to the experimental data.

The lattice parameters of the Pca21 phase are determined, and the enthalpies of different phases are compared under various pressures. The results indicate that the Pca21 phase has a much lower enthalpy than the other structures within the pressure range tested (Fig. 4). The lattice parameters of the Pca21 $TiO_2$ under 43 GPa are $a$=4.84 Å, $b$=4.51 Å, and $c$=4.55 Å. In the Pca21 $TiO_2$ phase, all identical Ti atoms occupy the 4a (0.5428, 0.7265, and 0.2112) sites, and all non-identical O atoms occupy the 4a (0.2477, 0.5629, and 0.4608) and 4a (0.3893, 0.0933, and 0.2989) sites. The hypothesis that Pca21 $TiO_2$ could revert directly to the columbite phase under decompression to -1 GPa is also validated in this work. From the trajectory file, a similar unusual transition appears at the $34^{th}$ step under a pressure of -1 GPa, indicated by squares in Fig. 1. The transition pressure (43 GPa) from the columbite to the Pca21 $TiO_2$ predicted by the *ab initio* calculations during the compression is different from the experimental value (48 GPa) [18]. The transition pressure (-1 GPa) from the Pca21 $TiO_2$ to the columbite $TiO_2$ predicted by the *ab initio* calculation during the decompression is also different from the experimental value (9 GPa) [18]. The difference in transition pressures between the theoretical and experimental calculations may be attributed to the fact that the *ab initio* calculations are performed in the ground state at zero temperature. The phonon dispersions of Pca21 $TiO_2$ are also calculated at 0, 10, 15, and 50 GPa. The results indicate that the Pca21 phase is stable between 15 and 50 GPa because no imaginary frequencies in the phonon spectra exist in this pressure range (the details can be seen in the EPAPS in Ref. 24). This may be why cubic $TiO_2$ can exist in pressures ranging from 9 to 48 GPa in the experimental synthesis [18].

Even when the same pressure-transmitting medium (NaCl) is used to measure the bulk moduli, the measured values show discrepancies of about 20% for the columbite, 40% for the baddeleyite, 29% for the orthorhombic I, and 32% for the cotunnite phases [25]. It is thus unsurprising that discrepancies exist between the theoretical and experimental bulk moduli obtained for the system under study. The underlying physics behind such a discrepancy is not clear at present, considering that many possible factors, including the quality of different samples and the different methods employed for measuring bulk modulus, exist. In addition, working with data such that they fit the third-order Birch-Murnaghan equation of state may yield discrepancies [27]. The third-order Birch-Murnaghan equation of state may be written as:

$$P=\frac{3B_0}{2}[(\frac{V_0}{V})^{\frac{7}{3}}-(\frac{V_0}{V})^{\frac{5}{3}}]\{1-\frac{3(4-B^{'})}{4}[(\frac{V_0}{V})^{\frac{2}{3}}-1]\}, \qquad (1)$$

where $V$ and $V_0$ are the volumes at pressure $P$ and the equilibrium volume at ambient pressure, respectively; and $B_0$ and $B'$ are the bulk modulus at ambient pressure and its pressure derivative, respectively. The uncertainty of the positions of the O atoms gives rise to large discrepancies in the bulk modules of the $TiO_2$ polymorphs. Many publications have reported that the pressure derivative B' is ~ 4.0 [6, 20, 25]. Using the squared residuals fitting method and choosing B' as the adjustable parameter, Hamane *et al*. found that smaller B' values result in larger $B_0$ values. The optimal value for cotunnite $TiO_2$ is B' = 4.25. Thus, this result is expected to be helpful in determining the same values for the other $TiO_2$ polymorphs. Table I lists our calculated results and compares them with reports in Refs. [18-20, 26]. The table shows that lower B' values result in higher $B_0$ values. The local density approximation method leads to overestimated $B_0$ values for the $TiO_2$ polymorphs because it underestimates $V_0$. However, our calculated $V_0$ (115.5 Å$^3$) and $B_0$ (207 GPa) for Pca21 $TiO_2$ are in excellent agreement with the experimental data (115.5 Å$^3$, 202 ± 5 GPa) [18]. In addition, the calculated value of B' (4.24) is consistent with the value of B' (4.25), as predicted in Ref. [25].

The calculated volume-pressure curves of the three possible phases (pyrite, fluorite, and Pca21) are shown in the inset of Fig. 4(a). They reveal that fluorite $TiO_2$ is the most incompressible phase among all the predicted phases, while Pca21 $TiO_2$ is more compressible than the fluorite and pyrite $TiO_2$ phases. Swamy and Muddle [19] indicated that the calculated values of $B_0$ for the pyrite and fluorite phases were significantly larger than the experimental values because of the coexistence of many possible phases in the synthesized sample [18]. Combined with the simulated XRD patterns and the equation of state, we provide direct evidence from the atomic level that the distortions of the O atoms play a dominant role in defining the compressive property of the sample. For the fluorite phase at the transition pressure of 43 GPa, Ti-O bonds with bond lengths of 2.01 Å have a coordination number of eight. In contrast, for Pca21 $TiO_2$, at the transition pressure of 43 GPa, the Ti-O bonds with average bond lengths of 1.966 Å (bond lengths ranged from 1.86 to 2.07 Å) have a coordination number of seven. Due to the very small volume difference between the fluorite (99.8 Å$^3$) and Pca21 (99.5 Å$^3$) $TiO_2$ phases at 43 GPa, the bonding instability in the Pca21 phase leads to a significant degree of bond-length fluctuations, which may decrease the coordination number of the Ti-O bonds. Based on Cohen's empirical formula: $B_0 \alpha AN_c/d^{3.5}$, where A is a constant, $N_c$ is the coordination number of a chemical bond, and d is the bond length [28], the decrease in the coordination number of the chemical bonds in Pca21 $TiO_2$ with respect to fluorite $TiO_2$ is one of the reasons for the large reduction of bulk modulus in cubic $TiO_2$ polymorphs. As such, Pca21 $TiO_2$ is more compressible than the fluorite phase within the pressure range under study [Fig. 4(a)]. The minute distortions of the O atoms dominate the unexpected reduction (~34%) in the bulk modulus at pressures of 277 GPa for the fluorite $TiO_2$, 207 GPa for the Pca21 $TiO_2$, and 202 GPa for the measured value in high-pressure cubic phases. We believe that this evidence clarifies the ambiguity of the bulk modulus in the high pressure phases of $TiO_2$. For example, if there exists a similar modified cotunnite phase, the expected bulk modulus reduction of ~30% with respect to cotunnite $TiO_2$ (431 GPa) [1] is in good agreement with independent experimental values of 312 ± 34 GPa [6] and 294 ± 9 GPa [25].

In conclusion, using *ab initio* calculations, we showed that an unusual and abrupt change in the energy curve of columbite $TiO_2$ at ~43 GPa produces modified fluorite $TiO_2$, a structure that had been theoretically conceived but never confirmed. The modified fluorite $TiO_2$ showed improved simulated XRD patterns and reversed to columbite $TiO_2$ under decompression to about -1 GPa.In particular, tiny distortions of the O atom positions result in an unexpected reduction in bulk modulus of about 34% in its high-pressure cubic phases. All of these are in good agreement with the experimental results. This is a good investigative study on the compressive properties of such group of homologous materials.

This work is supported by the National Basic Research Program of China under Grant No. 2006CB921805 and the Postdoctoral Fund of China under Grant No. 20090460685.

Table I. Zero-pressure bulk modulus and related properties of TiO2 polymorphs.

| Phase | Method | $V_0$ (Å$^3$) | $B_0$ (GPa) | $B'$ | Reference |
|---|---|---|---|---|---|
| Fluorite | VASP-GGA | 112.70 | 277 | 4.07 | This work |
| | CRYSTAL-GGA | 112.75 | 395 | 1.75 | [19] |
| | B3LYP | 112.13 | 390 | 2.06 | [19] |
| | BSTATE-GGA | 112.11 | 272 | 4.66 | [20] |
| | VASP-LDA | 107.08 | 309 | 4.46 | [20] |
| Pyrite | VASP-GGA | 117.73 | 258 | 4.27 | This work |
| | CRYSTAL-GGA | 118.62 | 220 | 4.86 | [19] |
| | B3LYP | 117.26 | 258 | 4.35 | [19] |
| | BSTATE-GGA | 116.65 | 272 | 4.58 | [20] |
| | VASP-LDA | 112.10 | 298 | 4.15 | [20] |
| Pca21 | VASP-GGA | 115.46 | 207 | 4.24 | This work |
| | Experiment | 115.50 | 202 | 1.3 | [18] |
| Rutile | VASP-GGA | 64.34 | 221 | 4.8 | This work |
| | CRYSTAL-GGA | 63.78 | 215 | 5.35 | [19] |
| | B3LYP | 63.42 | 224 | 5.64 | [19] |
| | Experiment | 62.44 | 211 | 6.76 | [26] |

FIG. 1. Relationships between the total energies and optimization steps of columbite $TiO_2$ at the transition pressure of 43 GPa (circles) and of Pca21 $TiO_2$ at the transition pressure of -1 GPa (squares). An unusual energy jump is observed at the 30$^{th}$ step of the columbite at 43 GPa. This jump is similar to the abnormal transition of Pca21 $TiO_2$ at the 34$^{th}$ step. Finally, the columbite transfers to Pca21$TiO_2$ at the 41$^{st}$ step, while Pca21 $TiO_2$ retransfers to the columbite at the 47$^{th}$ step.

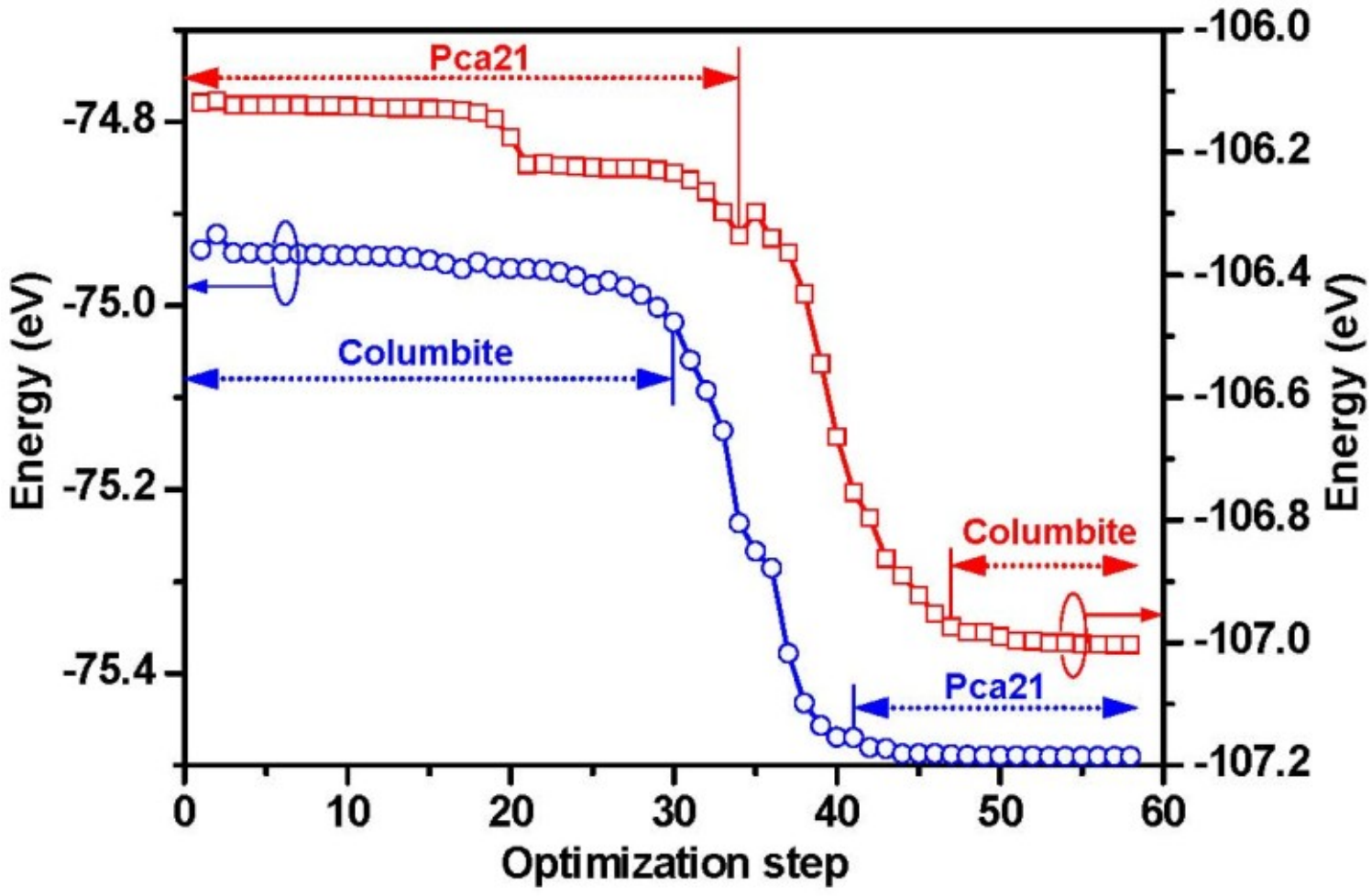

FIG. 2. Projections along the [010] direction of a 2×2×2 columbite $TiO_2$ supercell at 43 GPa. The O and Ti atoms are represented by small and large circles, respectively. (a), (b), and (c) show snapshots of the optimized 1st, 30$^{th}$, and last steps. (d) shows the fluorite $TiO_2$ for comparison.

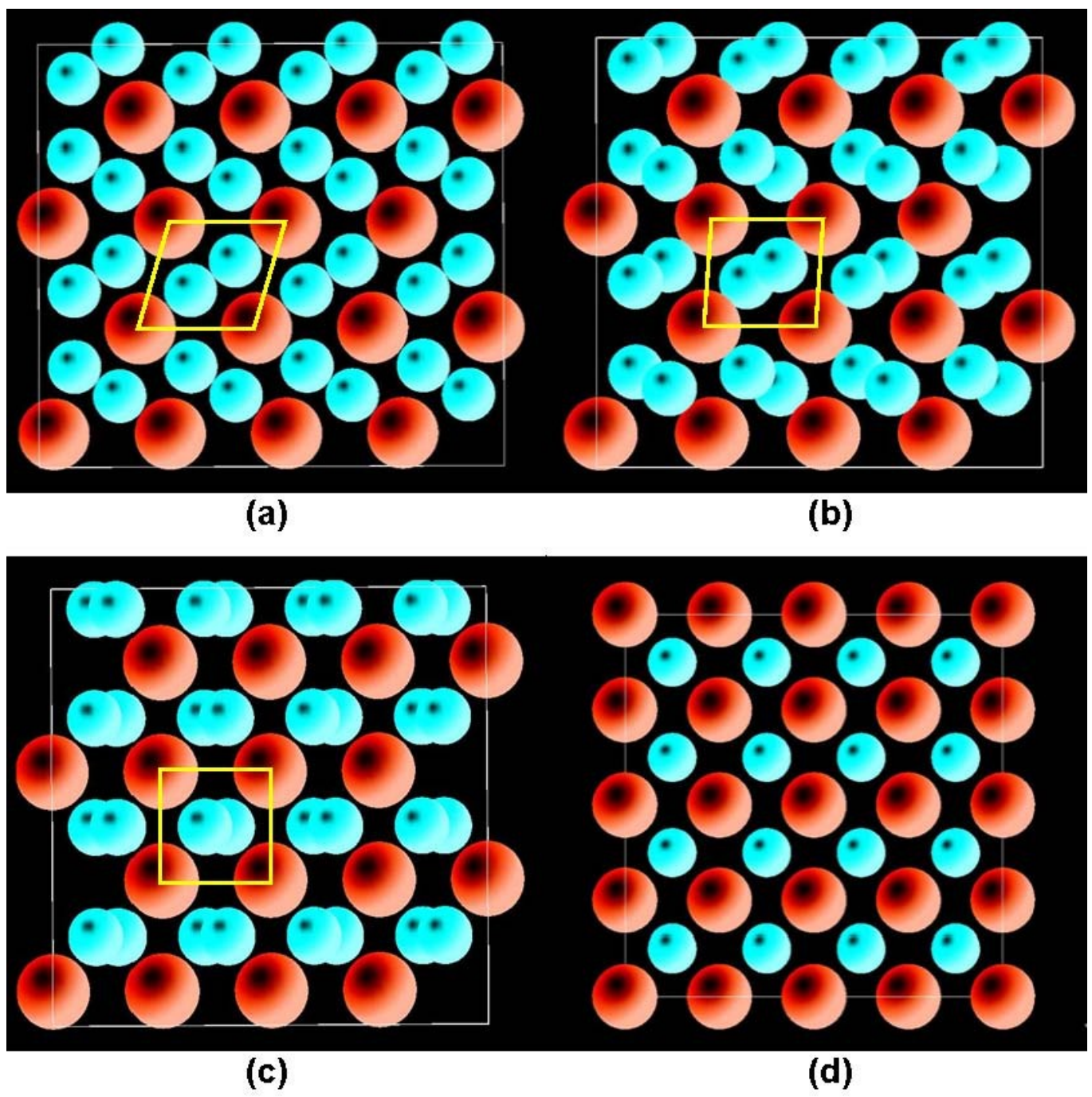

FIG. 3. Simulated XRD patterns of the pyrite, fluorite, and Pca21 structures at 0.6996 Å and 43 GPa in comparison with the experimental results at 48 GPa. The distinct 102 peak reflects the structure of cotunnite (O II).

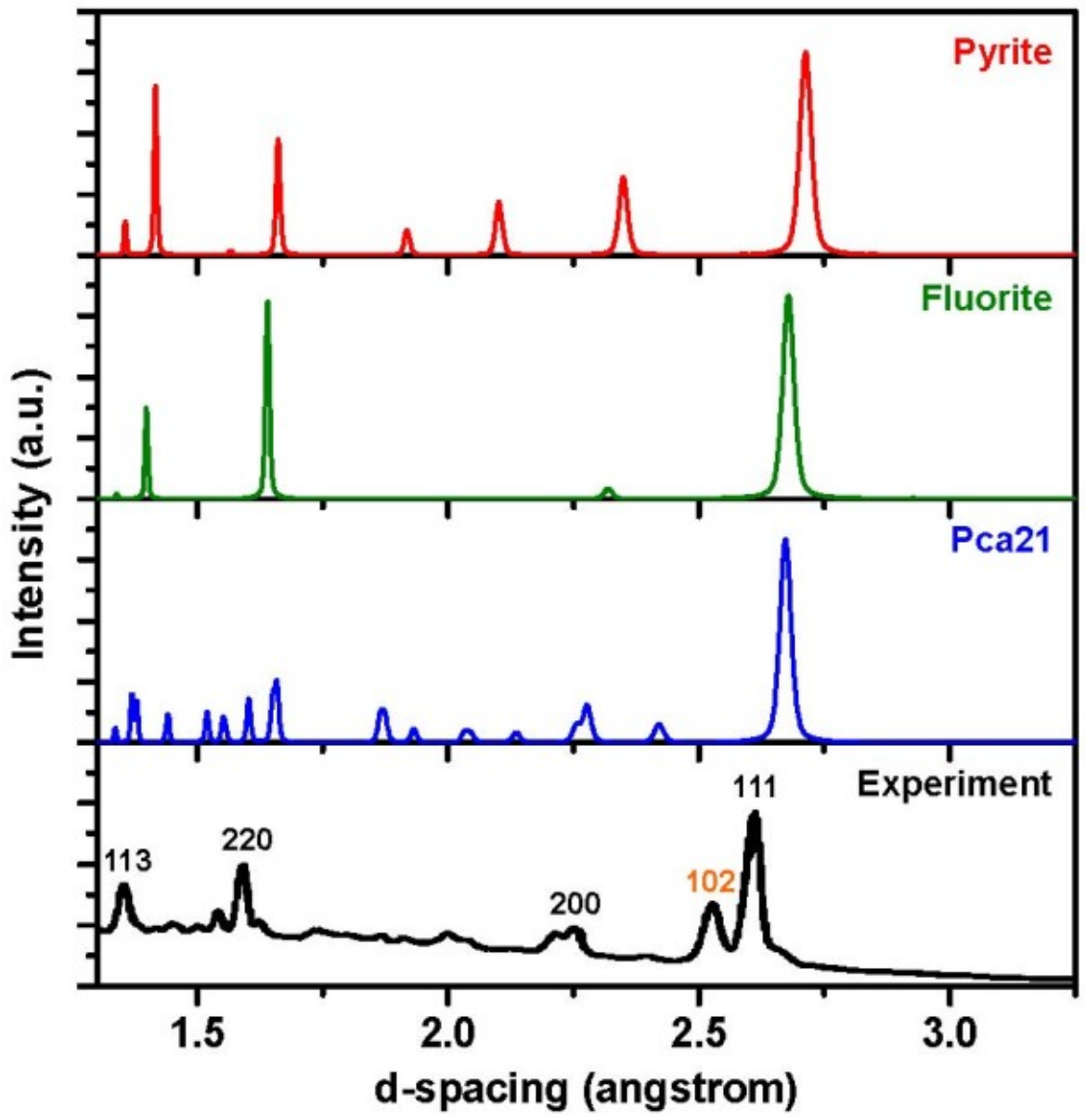

FIG. 4. Enthalpies of the pyrite, fluorite, and Pca21$TiO_2$ polymorphs in the pressure range of 0-50 GPa. The enthalpy difference is based on that of anatase $TiO_2$. The inset (a) shows the pressure-volume relations of the pyrite, fluorite, and Pca21 $TiO_2$ polymorphs. The inset (b) shows the fitting of the third-order Birch-Murnaghan equation of state with Pca21 $TiO_2$.

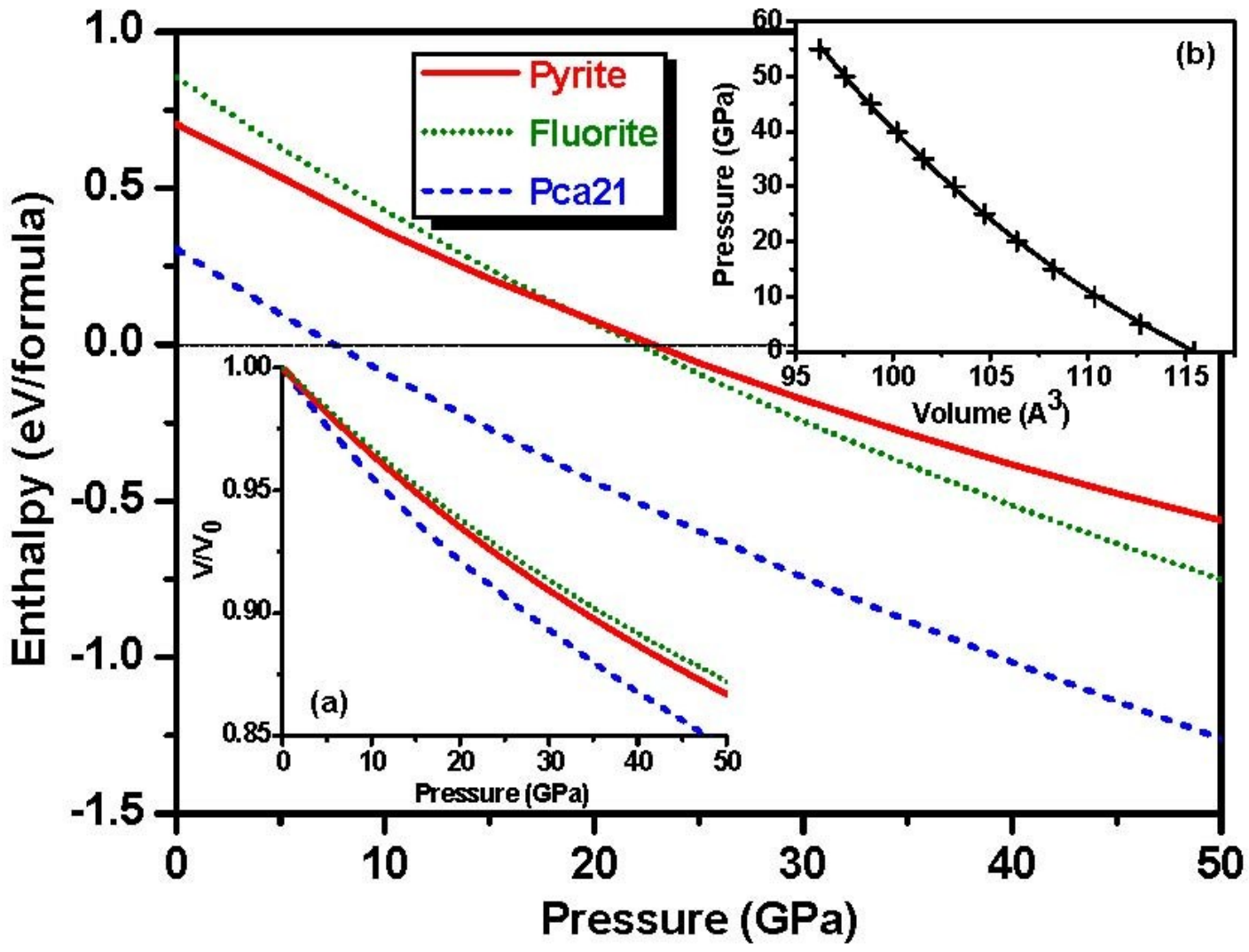